\documentclass[preprintnumbers,amsmath,amssymb]{revtex4}

\usepackage{dcolumn}
\usepackage{bm}
\usepackage{tabularx} %
\usepackage{color}

   \usepackage{graphicx}

\newcommand{\PreserveBackSlash}[1]{\let\temp=\\#1\let\\=\temp}

\begin{document}


\title{An approximate method for calculating transfer integrals based on the ZINDO Hamiltonian}

\author{James Kirkpatrick}
        \email{james.kirkpatrick@ic.ac.uk}
\affiliation{Centre for Electronic Materials and Devices, Department of Physics, Imperial College London, Prince Consort Road, London SW7 2BW, U.K.}

\date{\today}

\begin{abstract}
In this paper we discuss a method for calculating transfer integrals based on the ZINDO Hamiltonian which requires only a single self consistent field on an isolated molecule to be performed in order to determine the transfer integral for a pair of molecules. This method is compared to results obtained by projection of the pair of molecules' molecular orbitals onto the vector space defined by the molecular orbitals of each isolated molecule. The two methods are found to be in good agreement using three compounds as model systems: pentacene, ethylene and hexabenzocoronene.
\end{abstract}

\pacs{Valid PACS appear here}
\maketitle

\section{Introduction}

In a disordered material - such as a glass of small molecules or a conjugated polymer film- charge transport can be modeled as a series of discrete hops on an idealized lattice; rates are controlled by parameters distributed according to some empirical distribution chosen to fit the experimentally measured field and temperature dependence of mobility\cite{Bassler_91, Yu_01, Novikov_95} . The fundamental mechanism underpinning charge transport in many disordered organic solids is thought to be small polaron hopping, which - in the high temperature limit - can be described by the Marcus Equation \cite{Marcus_85}:

\begin{equation}
\frac{|J|^2}{\hbar} \sqrt{\frac{\pi}{\lambda kT}} exp(-\frac{(\Delta E -\lambda)^2}{4 \lambda kT} )
\end{equation}

where $ J $ represents the transfer integral, $ \Delta E $ represents the difference in site energies, $ \lambda $ is the reorganization energy and all other symbols have their usual meanings. In our opinion, it would be a significant improvement if the parameters of this equation could be calculated for realistic morphologies, helping to clarify the relationship between chemical structure and charge mobility and reducing the number of free parameters in the modeling of data: the difficulty lies in the fact that simulation volumes can contain millions of molecules and therefore these parameters must be calculated using efficient, fast algorithms. In this paper we wish to discuss a computational prerequisite to solving the dynamics of electron motion in a disordered medium: the design of efficient algorithms for computing the transfer integral $ J $.

The method is based on the ZINDO hamiltonian \cite{Zerner_73} and makes some approximations to allow calculation of transfer integrals without the necessity for performing self consistent field (SCF) calculations on pairs of molecules: only a single SCF calculation on 1 isolated molecule will be performed. It will require only the calculation of atomic overlap and can be thought of as being based on the calculation of molecular orbital overlap: for this reason we dub the method Molecular Orbital Overlap (MOO). The ZINDO Hamiltonian has been used extensively to calculate transfer integrals \cite{Bredas_04}, even though recently it has become apparent the simply taking the splitting of the top two molecular orbitals is not always accurate because of polarization effects; another method which has been used by Siebbeles and coworkers exploits the molecular fragment capabilities of ADF to calculate $ J $ \cite{Siebbeles_03}, although it has been pointed out by Valeev and co-workers that this ought to be corrected for molecular overlap \cite{Filho_06}. In this paper we will show how to rewrite the Fock matrix from the ZINDO method in terms of localized monomer orbitals  by orbital projection to obtain results similar to those obtained by the molecular fragment method. We will compare the results with those from MOO, showing how the agreement is very good. The model systems we study are ethylene, pentacene and hexa-benzocoronene (HBC). It should also be noted that we have carried out these test only for pairs of identical molecules and that the following derivations are labelled accordingly; extension to the general case is trivial.

\section{Method}

The definition of the transfer integral for charge transport from molecule A to molecule B is: 

\begin{equation}
J = < \mathbf{\Phi^A} | H | \mathbf{\Phi^B} > 
\label{Jeq0}
\end{equation}
where $ H $ represents the Hamiltonian for the system, $ \mathbf{\Phi} $ represents the multi-electron wavefunction of the molecule and the labels $ A $ and $ B $ denote whether the charge is localized on molecule A or B. Assume the multi-electron wavefunctions are described by single Slater determinants and invoke the frozen orbital approximation to argue the $ \mathbf{\Phi}^A $ and $ \mathbf{\Phi}^B $ differ only by the highest occupied molecular orbital (HOMO) on molecule A and molecule B, which will be singly occupied in molecule A and B respectively. If we were interested in transport of negative charge, we would obviously use the lowest unoccupied molecular orbital (LUMO). Using Slater Rules \cite{Szabo} we can evaluate the previous equation as: 
\begin{equation}
J=<\phi_{homo^A} | F | \phi_{homo^B}>
\label{J_eq}
\end{equation}
where $ F $ represents the Fock matrix and $ \phi_{homo^A} $ and $\phi_{homo^B} $ represent the HOMOs of molecule A and B respectively. We will always consider the case of calculating transfer integrals for two Slater determinants which differ in one molecular orbital only, therefore evaluating equation \ref{J_eq} is always going to be our task.

\subsection{Projective Method}

In order to solve equation \ref{J_eq} we will invoke the spectral theorem, project the molecular orbitals (MOs) of the dimer onto a basis set defined by the MOs of the individual molecules, then  - knowing the eigenvalues of the MOs of the dimer - we will reconstruct the Fock matrix in the basis set of the MOs of individual moleculs and simply read off $ J $ from the appropriate indeces of our new Fock matrix. The basis set defined by the MOs of the individual molecules $ C_{loc} $ will be defined as:

\begin{equation}
C_{loc}=
\begin{array}{cccccccc}
\phi^1_1 & \phi^2_1 & ... & \phi^{\frac{N}{2}}_1 & 0 & 0 & 0 & 0 \\
\phi^1_2 & \phi^2_2 & ... & \phi^{\frac{N}{2}}_2 & 0 & 0 & 0 & 0 \\
...      & ...      & ... & ...          & 0 & 0 & 0 & 0 \\
\phi^1_{\frac{N}{2}} & \phi^2_{\frac{N}{2}} & ... & \phi^{\frac{N}{2}}_{\frac{N}{2}} & 0 & 0 & 0 & 0 \\
0 & 0 & 0 & 0 & \phi^{\frac{N}{2}+1}_{\frac{N}{2}+1} & \phi^{\frac{N}{2}+2}_{\frac{N}{2}+1} & ... & \phi^{N}_{\frac{N}{2}+1} \\
0 & 0 & 0 & 0 & \phi^{\frac{N}{2}+1}_{\frac{N}{2}+2} & \phi^{\frac{N}{2}+2}_{\frac{N}{2}+2} & ... & \phi^{N}_{\frac{N}{2}+2} \\
0 & 0 & 0 & 0 & ... & ... & ... & ... \\
0 & 0 & 0 & 0 & \phi^{\frac{N}{2}+1}_{N} & \phi^{\frac{N}{2}+2}_{N} & ... & \phi^{N}_{N} 
\end{array}
\end{equation}
where $ \phi^i_j $ labels the component in terms of atomic orbital (AO) $ j $ of molecular orbitals $ i $. The AOs are numbered so that the first $ \frac{N}{2} $ orbitals are localized on molecule A, and the second $ \frac{N}{2} $ are localized on molecule B, similarly the molecular orbitals localized on molecule A are labeled with the first $ \frac{N}{2} $ labels and the ones on molecule B are labeled with the second $ \frac{N}{2} $ labels. The localized orbitals are deduced from SCF calculations on the isolated molecules respectively, seeing as how the two molecules are identical only one SCF calculation is required, the other set of orbitals can be obtained by rotating the orbitals of the first molecule according to the spatial orientation of the second.

In order to project the MOs of the dimer $ C_{dym} $ onto $ C_{loc} $ and obtain the orbitals of the dimer in the localized MO basis set all we have to do is invoke the spectral theorem and obtain: 

\begin{equation}
C_{dym}^{loc} = C_{loc}^t C_{dym}
\end{equation}

where the superscript $ ^t $ denotes transposition and $ C_{dym}^{loc} $ represents the orbitals of the dimer in the localized basis set. All that is left to do is use the dimer eigenvalues $ \epsilon_{dym} $ and rewrite the Fock matrix $ F $ in the new basis set to obtain the Fock matrix in the localized basis set:
\begin{equation}
F^{loc} = C_{dym}^{loc\;t} \epsilon_{dym} C_{dym}^{loc}
\end{equation}
where  $ F^{loc} $ represents the Fock matrix in the localized basis set and the eigenvalues $ \epsilon_{dym} $ have been written in diagonal matrix form. Now transfer integrals can simply be read from the off-diagonal elements of this matrix, it we are interested in the transfer integral between the HOMO on molecule A and the HOMO on molecule B - and assuming that the HOMO is the $ i^{th} $ orbital of molecule A - we would simply read the element $ F^{loc}_{i, i+\frac{N}{2}} $.

\subsection{Molecular Orbital Overlap calculation of $ J $}

In this section we will explain how to evaluate directly equation \ref{J_eq} in the ZINDO approximation and how further approximation can be used to make the running of a SCF calculation on a dimer unnecessary. If we write the HOMOs on molecule A and B, labelling the AOs in the same fashion we used in the definition of $ C_{loc} $, we can see that the only elements of the Fock matrix which we need to calculate are the off-diagonal blocks of the Fock matrix connecting the AOs on molecule A with the AOs on molecule B. These elements will necesseraly involve AOs on different centres, therefore their form will be: 

\begin{equation}
F_{\mu \nu} = \bar{S}_{\mu \nu} \frac{( \beta_{A} + \beta_B ) }{2} + P_{\mu \nu} \frac{\gamma_{AB}}{2} 
\end{equation}
where $ \bar{S} $ represents the overlap matrix of atomic overlaps with $ \sigma $ and $ \pi $ overlap between $ p $ orbitals weighed differently, $ A $ and $ B $ labels the two atomic centres that the $ \mu $ and $ \nu $ atomic orbitals are centred on, $ \beta_A $ labels the ionization potential of molecule $ A $ , $ P_{\mu \nu} $ labels the density matrix and finally $ \gamma $ is the Mataga-Nashimoto potential. We assume that $ P_{\mu \nu} $ is block diagonal and therefore does not contribute to the elements of the Fock matrix we are interested in calculating. This assumption for the dimer orbitals will hold both if the dimer orbitals are identical to the monomer ones or if each pair of orbitals of the dimer can be written as a constructive/destructive combination of pairs of monomer orbitals. To see why the latter is the case consider two particular dimer occupied orbitals $ \phi_{i} $ and $\phi_{i+1} $ which are formed from a bonding and anti-bonding combination of the occupied monomer orbitals $ \phi_{Aj} $ and $ \phi_{Bj} $. The contribution of these two orbitals to the density matrix will be of the form: 

\begin{subequations}
\begin{equation}
P^{i,i+1} = 2.0 (\phi_{i}^{t} \phi_i + \phi_{i+1}^{t} \phi_{i+1})
\end{equation}
\begin{equation}
= 2.0 ( (\phi_{Aj} + \phi_{Bj})^t) (\phi_{Aj} + \phi_{Bj}) + 2.0 (\phi_{Aj} - \phi_{Bj})^{t} (\phi_{Aj} - \phi_{Bj})  )
\end{equation}
\begin{equation}
= 2.0 ( 2 \phi_{Aj}^{t} \phi_{Aj} + 2 \phi_{Bj}^{t} \phi_{Bj} )
\end{equation}
\end{subequations}

Where $ P^{i,i+1} $ represents the contribution of dimer orbitals $ \phi_i $ and $ \phi_{i+1} $ to the density matrix. Because all monomer orbitals $ \phi_A $ and $ \phi_B $ are localized on one molecule only, this contribution will be block diagonal, also because all contributions to the density matrix will be of this form, the density matrix will be - overall  - block diagonal. The task of determining values for the Fock matrix has therefore been reduced to the comparatively simple task of determining $ \bar{S} $, the weighed atomic orbital overlap. Atomic overlaps between 1s, 2s and 2p orbitals can be determined analytically using the expressions derived in \cite{Mulliken_49}, the $ \pi $ and $ \sigma $ components of the $<p|p>$ overlaps must be weighed according to the appropriate proportionality factors, in accordance with the scheme devised by Zerner and coworkers. This can be done without the need to perform a SCF calculation on the dimer, thereby achieving our set goal of estimating transfer integrals for dimers whilst performing only one calculation on the monomers to obtain the orbitals $ \phi_{homo^A} $ and $\phi_{homo^B} $.

\section{Computation Details}

For both the projective and MOO methods, some information has to be extracted from a SCF calculation: in the case of the projective method we need the monomer MOs, the dimer MO and the dimer eigenvalues. For the MOO method, all we need are the monomer MOs for which we want to evaluate the expectation of the Fock matrix for. All information from self consistent field calculations is extracted from g03 \cite{Gaussian}. The matrix operations for the projective method and the analytic solution to the AO overlaps for the MOO method are all computed with in-house code. The MOO libraries are written for row 1 and 2 atoms and will soon be realised on Gnu Public Licence. Both methods require starting geometries for each monomer, these were computed with g03 and the the B3LYP/6-31g* level.

\section{Results}

\begin{figure}
\includegraphics[width=10cm]{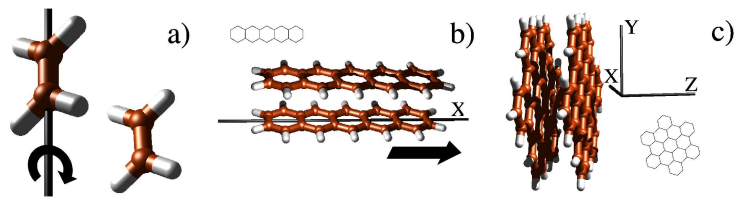}
\caption{The three molecules and their orientations used to compare the projective and MOO methods for calculating $ J$. a) Shows an ethylene dimer at a distance of 5 \AA \ and the axis around which one of the two ethylenes is rotates. b) Shows a pentacene dimer at a distance of 3.5 \AA \ and the axis along which one of the two molecules is shifted. Finally c) shows a pair of HBC molecules and the three axis used to displace or rotate the two molecules. b) and c) contain insets showing stick models of the molecules under consideration with all hydrogens removed.}\label{key}
\end{figure}

In this section we will show the results of comparison of the results from the projective and MOO methods. We will use ethylene and pentacene as examples of a conjugated molecule and HBC as an example of a high symmetry conjugated molecule where we will show that it is necessary to calculate different transfer integrals to define an effective transfer integral. The geometries which we use to compare these methods are shown in figure \ref{key}, these are: rotation around the C=C bond for one of the two molecules in an ethylene dimer, slip along the long axis of one of two pentacene molecules and x,y,z displacement of a HBC molecule in a dimer. 

\begin{figure}
\includegraphics[angle=270]{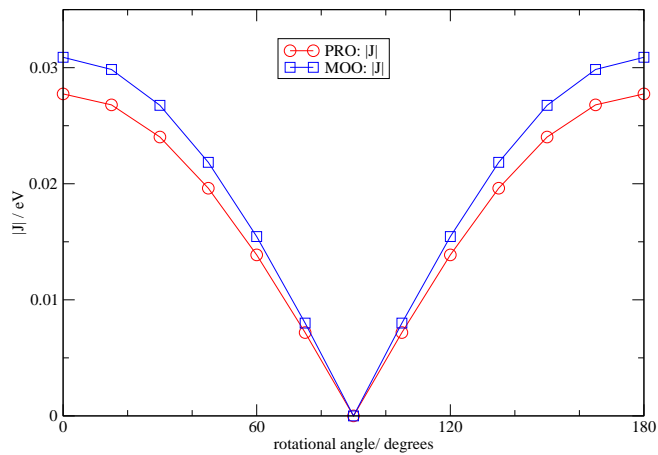}
\caption{ $|J|$ as a function of rotation around the C=C double bond of one of the ethylenes calculated using the MOO and projective (PRO) methods. The two ethylene molecules are at a centre to centre distance of 5 \AA \ .}\label{ethylene}
\end{figure}

\begin{figure}
\includegraphics[angle=270]{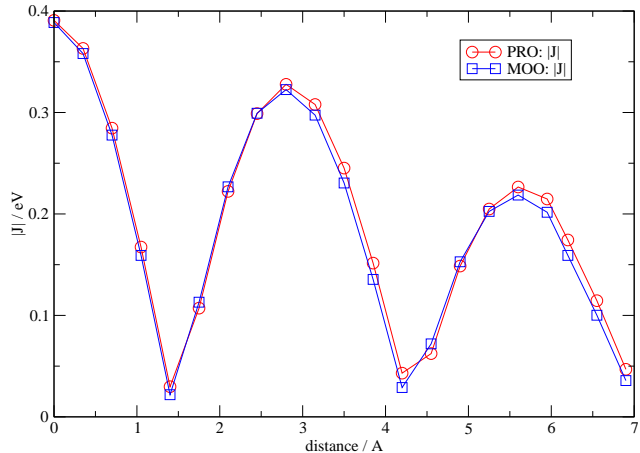}
\caption{$|J|$ as a function of slip along the x axis for a pair of pentacene molecules calculated using the MOO and projective (PRO) methods. The distance between the $ \pi $ planes of the two pentacene molecules is 3.5 \AA \ .}\label{pentacene}
\end{figure}

The results for ethylene are shown in figure \ref{ethylene}, as it is expected from the planarity of the molecule, the transfer integral falls to 0 for perpendicular. These results are in qualitative agreement with the DFT results of Valeev and co-workers \cite{Filho_06}, even though the value of the transfer integral from ZINDO is roughly half of that from DFT. Certainly the projective and MOO methods are in excellent agreement, with a discrepancy between the two methods of approximately 10\%. The results for pentacene are shown in figure \ref{pentacene} and -again- are consistent with each other.

\begin{figure}
\includegraphics[angle=270]{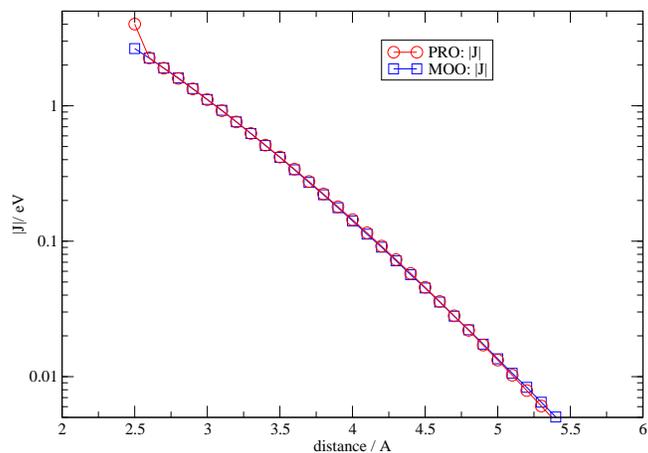}
\caption{ $|J|$ as a function of z displacement for two HBC molecules in the same orientation calculated using the MOO and projective (PRO) methods.}\label{HBC_z}
\end{figure}

Before considering the case of HBC, let us make a few comments on how to approach the problem of determining transfer integrals for molecules with symmetry induced degeneracy of the frontier orbitals. The physical phenomenone which one would expect to occur in such a situtation is that - upon charging - the molecule will lose its symmetry by Jahn-Teller distorsion and that therefore charge transfer will occure between non-degenerate orbitals. In order to avoid having to calculate many different transfer orbitals for the different possible distorsions, an approach which has been used in the literature - section IV D of \cite{Newton_91} - is to simply take the root means square value of the four possible integrals between the two degenerate orbitals, which in our case would be $ J_{homo^A \; homo^B} $, $ J_{homo^A \; homo-1^B}$ , $ J_{homo-1^A \; homo^B}$ and $ J_{homo-1^A \; homo-1^B} $. Let us justify this approximation by generalizing equation \ref{J_eq}. Assume that $ \mathbf{\Phi^A} $ and $ \mathbf{\Phi^B} $ are linear combination two Slater determinant, each corresponding to either the HOMO or the HOMO-1 of either molecule being simply occupied. Label these two Slater determinants $ \mathbf{\Phi^{A1}} $ and $ \mathbf{\Phi^{A2}} $ respectively for molecule A and similarly for molecule B. The linear combination $ \mathbf{\Phi^A} $ can then be written: 

\begin{equation} 
\mathbf{\Phi^A} = cos(\chi) \mathbf{\Phi^{A1}} + sin(\chi) \mathbf{\Phi^{A2}}
\label{Slat_deg}
\end{equation}
where $\chi$ represents the mixing angle for the two configurations. A similar equation can be written for the localized state on molecule B. When equation \ref{Jeq0} is estimated for the form of Slater determinants from \ref{Slat_deg} one obtains form of \ref{J_eq} which involve the mixing angles for molecule A and molecule B and 4 expectation values for the Fock matrix: $ J_{homo^A \; homo^B} $, $ J_{homo^A \; homo-1^B}$ , $ J_{homo-1^A \; homo^B}$ and $ J_{homo-1^A \; homo-1^B} $. If one squares this expression to obtain the form of the observable $ |J|^2 $ and averages this equation over the two mixing angles, one obtains an expression for the effective transfer integral $ |J_{eff}|^2 $  as the average of the other 4 transfer integrals squared. In the case of z displacement, $ J_{homo-1^A \; homo^B}$ and $ J_{homo^A \; homo-1^B}$ are both 0, and the other two terms are the same, in this case, we will plot $ J_{homo^A \; homo^B} $ as a function of distance. This quantity would be the same as half the splitting between the top 2 MOs and the next 2 MOs of a dimer and is equivalent to $ J_{eff} \sqrt(2) $. A plot of this transfer integral calculated using MOO and deduced from a ZINDO calculation with the projective method is shown in picture \ref{HBC_z}. Again it can be seen that the two methods are in very close agreement, with the exception of the pair of molecules at 2.5 \AA \ , we postulate that at this distance the assumption of block diagonal density matrix brakes down. The value obtained for this geometry can also be compared to some from the literature: in \cite{Lemaur_04} a quantitavely similar curve is reported from the splitting of the frontier orbitals of HBC.

\begin{figure}
\includegraphics[width=10cm]{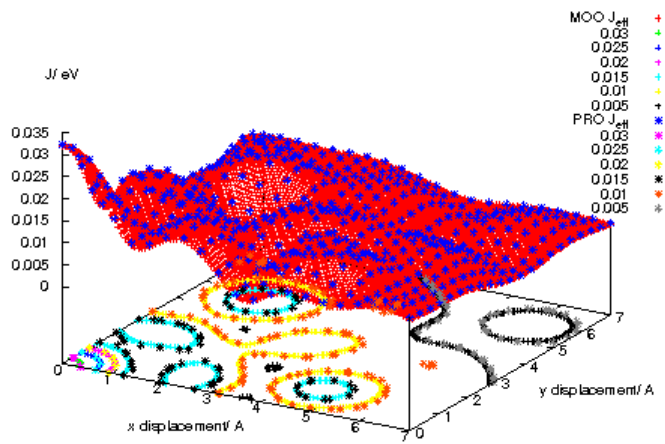}
\caption{ $J_{eff}$ as defined in the text for HBC calculated using the MOO and projective (PRO) methods as a function of x and y displacement. On the bottom are projected the contours of $J_{eff}$. }\label{HBC_xy}
\end{figure}

If the dimer is displaced in the xy direction, the terms $ J_{homo-1^A \; homo^B}$ and $ J_{homo^A \; homo-1^B}$  are no longer 0 and $ J_{homo^A \; homo^B} $ is no longer equal to $ J_{homo-1^A \; homo-1^B} $, in this case we will plot $J_{eff}$ as calculated from the transfer integrals using either the projective or the MOO methods. Figure \ref{HBC_xy} shows how - also for this case - the two methods are again in excellent agreement. 

\section{Conclusions}

We have shown how to use the spectral theorem to project orbitals of a dimer onto the localized basis set of MOs of the constituent monomers. We have argued that this method can be used to obtain results similar to those obtained by the method of fragment orbitals and have shown that in certain cases these results can be compared to those obtained by looking at the splitting of the frontier orbitals. We have shown that in all these cases the MOO method can be used to obtain results which are essentially the same as for the projective method, achieving our goal of determining transfer integrals whilst performing only 1 SCF calculation.

\clearpage

\begin{acknowledgments}
JK acknowledges the financial support of the UK Engineering and Physical Sciences (EPSRC) Research Council. We would also like to acknowledge the EPSRC National Centre for Computational Chemistry Software and all its staff for computer
time and support. J. Cornil and the whole of the Mons group are thanked for the useful and insightful conversation which sparked the ideas for this project.
\end{acknowledgments}

\end{document}